\documentclass[prl,aps,twocolumn,superscriptaddress]{revtex4}
\usepackage{graphicx,amssymb,amsmath,color,psfrag}
\usepackage{amsthm}
\usepackage{amsfonts}
\usepackage{algorithmic}
\usepackage{enumerate}
\usepackage{latexsym}
\begin{document}
\title{Electronic band gaps and transport properties in aperiodic bilayer graphene
superlattices of Thue-Morse sequence}

\author{Changan Li}
\affiliation{Department of Physics, Beijing Normal University,
Beijing 100875, China}

\author{Hemeng Cheng}
\affiliation{Department of Physics, Beijing Normal University,
Beijing 100875, China}
\author{Ruofan Chen}
\affiliation{Beijing Computational Science Research Center,
Beijing 100084, China}

\author{Tianxing Ma}
\email{txma@bnu.edu.cn}
\affiliation{Department of Physics, Beijing Normal University,
Beijing 100875, China}

\affiliation{Beijing Computational Science Research Center,
Beijing 100084, China}

\author{Li-Gang Wang}
\email{sxwlg@yahoo.com}
\affiliation{Department of Physics, Zhejiang University, Hangzhou 310027, China}

\affiliation{Beijing Computational Science Research Center,
Beijing 100084, China}
\author{Yun Song}
\affiliation{Department of Physics, Beijing Normal University,
Beijing 100875, China}

\author{Hai-Qing Lin}
\affiliation{Beijing Computational Science Research Center,
Beijing 100084, China}


\begin{abstract}
We investigate electronic band structure and transport properties in bilayer graphene superlattices of Thue-Morse sequence. It is interesting to find that the zero-$\overline{k}$ gap center is sensitive to interlayer coupling $t'$, and the centers of all gaps shift versus $t'$ at a linear way. Extra Dirac points may emerge at $k_{y}\ne$0, and when the extra Dirac points are generated in pairs, the electronic conductance obeys a diffusive law, and the Fano factor tends to be 1/3 as the order of Thue-Morse sequence increases. Our results provide a flexible and effective way to control the transport properties in graphene.
\end{abstract}

\maketitle



Graphene, a potential material with applications that continue to rise\cite{Novoselov2004,AHCastro2009,Peres2010,Chen2009,Ma2010,MZarenia2012,LBrey2009}.
The monolayer graphene is a semi-metal or zero-gap semiconductor,
and the bilayer graphene provides the first
semiconductor with a gap that can be tuned externally\cite{Zhang2009}.
In graphene, the low-energy charge carries behave as massless Dirac fermions near Dirac point (DP),
and the resulting linear energy dispersion relation leads to many interesting electronic and optical properties.
Recently, motivated by the experimental realization of graphene superlattices (GSLs) \cite{JC2008,ALVa2008},
enormous theoretical investigations have been done on the GSLs with periodic, qusi-periodic
and aperiodic electrostatic potentials or magnetic barriers \cite{Park2008,LGWang2010,PLzhao2011,TXMa2012,Liang2011,MK2011,ZhRzh2012,MBarb2010,XXGuo2011,Gong2012}.
It is well known that superlattices are vastly used to control the electronic band structure of
many conventional semiconducting materials and they have had huge impact on semiconductor physics \cite{Tsu2005,Cott1989}.
In  monolayer graphene supperlattice, it has been found that an unusual
DP appears in the band structure, and it is exactly located at the
energy which corresponds to the zero-$\overline{k}$ gap \cite{LGWang2010,TXMa2012}.
The location of the zero-$\overline{k}$ gap near DP is not only
independent on lattice constant but also insensitive to incident angles in
contrary with other Bragg gaps,
which results in the robust transport properties\cite{LGWang2010,XXGuo2011,PLzhao2011,TXMa2012}.

In this letter, we investigate the electronic band gaps and transport properties in bilayer graphene
superlattices (BLG SLs) of Thue-Morse (TM) sequence\cite{TM}.
As a typical aperiodic sequence, the TM lattice has been investigated
extensively \cite{NLiu1997,ZCheng1988,Luck1989,Jiang2005,Noh2011}.
It is a deterministically aperiodic structure and has a degree of order intermediate
between quasi-periodic and disordered system \cite{NLiu1997}. Contrary to the result of structure factor,
the electron behaviors of TM sequence show that it is more similar to a periodic system \cite{NLiu1997}.
It is known to have a continuous Fourier transform \cite{ZCheng1988} and
there is relationship between geometrical characters and their physical properties \cite{Luck1989}.
In the BLG SLs of TM sequence, it is very important to see that the zero-$\overline{k}$ gap can be tunable by
the interlayer coupling $t'$. The zero-$\overline{k}$ gap happens in the BLG SLs of TM sequence,
resulting in the robust transmission properties different from monolayer GSLs. Moreover, Extra DPs
arise and they are dependent on lattice constant.

We consider the electronic structure of BLG with energy and
wave vector close to the $K$ point, so the one-particle Hamiltonian for BLG
is
\begin{equation}
H=\left(
\begin{array}{cccc}
V(x)&\pi&t'&0\\
\pi^\dag&V(x)&0&0\\
t'&0&V(x)&\pi^\dag\\
0&0&\pi&V(x)%
\end{array}
\right).
\end{equation}
Here, $V(x)$ is the electrostatic potential which only depends on the coordinate $x$, $t'$ is the interlayer coupling, $\upsilon _{F}\approx 10^{6}$m/s is the Fermi velocity, $\pi$ and $\pi^\dag$ are the momentum operators \cite{MZarenia2012}, and $\pi=-i\hbar\upsilon _{F}\lbrack\frac{\partial }{\partial x}-i\frac{\partial}{\partial y}\rbrack$, $\pi^\dag=-i\hbar\upsilon _{F}\lbrack\frac{\partial }{\partial x}+i\frac{\partial}{\partial y}\rbrack $. The wave function is expressed by a four-component pseudospinors $\Phi={\left(\widetilde{\varphi}_{1},\widetilde{\varphi}_{2},\widetilde{\varphi}_{3},\widetilde{\varphi}_{4}\right)}^{T}$.
Due to the translation invariance in the $y$ direction, the wave function can be rewritten as $\widetilde{\varphi}_{m}=\varphi_{m}e^{ik_{y}y}$, m=1,2,3,4.
By solving the eigenequation, the wave functions at any two positions $x$ to $x+\Delta x$ inside the $j$th potentials
can be related by a transfer matrix
\begin{equation}
M_{j}=
\begin{pmatrix}
M_{+}&0\\
0&M_{-}
\end{pmatrix},
\end{equation}
and
\begin{equation}
M_{\pm}=
\begin{pmatrix}
\frac{\cos(q_{j}\Delta x\mp\Omega_{j})}{\cos\Omega_{j}}&i\frac{k_{j}}{q_{j}}\sin(q_{j}\Delta x)\\
i\frac{k_{j}'\sin(q_{j}\Delta x)}{k_{j}\cos\Omega_{j}}&\frac{\cos(q_{j}\Delta x\pm\Omega_{j})}{\cos\Omega_{j}})
\end{pmatrix}.
\end{equation}
where $k_{j}=(E-V_{j})/\hbar\upsilon _{F}$, $t'\to t'/\hbar\upsilon_{F}$,
$q_{j}=sign(k_{j})\sqrt{k_{j}^{2}-k_{y}^2-t'k_{j}}$, $k_{j}'^{2}=k_{y}^{2}+q_{j}^{2}$ and $\Omega_{j}=arcsin(k_{y}/k_{j}')$.
For an $n$-th TM sequence, $S(1)=AB$, $S(2)=ABBA$ and naturally $S(3)=ABBABAAB$, and so on.
Here, $A$($B$) is considered as alternating barrier $V_{A}$ ($V_{B}$) with its width $w_{A}$ ($w_{B}$).
After a long but straight algebraic deduction, we arrive at the transmission coefficient $t$=$t(E,k_{y})$, which can be expressed as
\begin{equation}
t=\frac{2(k'/k)\cos\Omega_{0}}
{(k'/k)(x_{22}e^{-i\Omega_{0}}+x_{11}e^{i\Omega_{e}})-(k'/k)^{2}x_{12}e^{i(\Omega_{e}-\Omega_{0})}-x_{21}}.
\end{equation}
$x_{i,j}(i,j=1,2)$ is the element of total transfer matrix \{X[$S(n)$]\}=$\prod_{j=1}^{N}M_{j}$, where $N$
is the total number of layers of GSL.
Because the transfer matrix X[$S(n)$] and $M_{j}$ are both unimodular, it is convenient to derive the recursion
relations of the trace map of the $n$-th BLG SLs TM sequence as
\begin{equation}
x_{n}=\frac{1}{4} Tr\{X[S(n)]\}=4x_{n-2}^{2}(x_{n-1}-1)+1.
\end{equation}

\begin{figure}[tbp]
\centering
\includegraphics[scale=0.45]{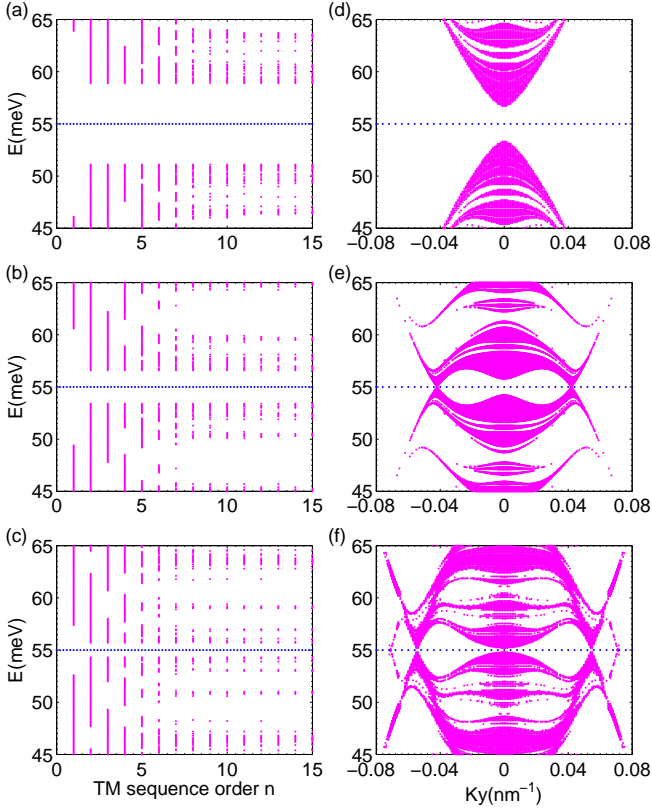}
\caption{(Color online)(a-c) Trace maps of TM sequence for (a) $w_{A}$=$w_{B}$=15 nm, (b) $w_{A}$=$w_{B}$=25 nm, and (c) $w_{A}$=$w_{B}$=60 nm, with  $k_{y}$=0.015 nm$^{-1}$. (d, e, f) are the corresponding electronic band structures to (a, b, c), respectively, with the order $n$=8. The other parameters are $V_{A}$=100 meV, $V_{B}$=0 meV, $t'$=10 meV in all cases.}
\label{Fig:fig1}
\end{figure}
\begin{figure}[tbp]
\includegraphics[scale=0.41]{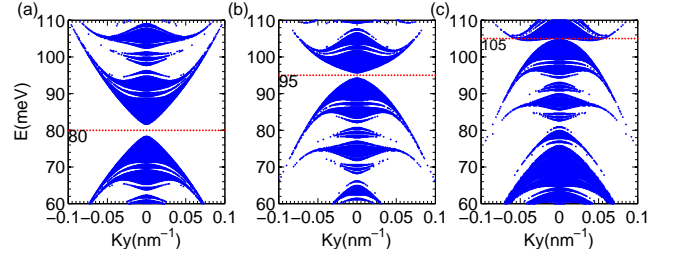}
\centering
\caption{(Color online)Electronic band structure for BLG SLs of TM sequence with $n$=8, (a) $w_{A}/w_{B}$=1, (b) $w_{A}/w_{B}$=3/2,
and (c) $w_{A}/w_{B}$=2 with $w_{B}$=10 nm, $V_{A}$=150 meV, $V_{B}$=0 meV, and $t'$=10 meV in all cases.}
\label{Fig:fig2}
\end{figure}

We plot the trace maps for BLG SLs of TM sequence with the order $n$ at $k_{y}$=0.015 nm$^{-1}$
in Figs. \ref{Fig:fig1}(a-c) and the corresponding band structures are plotted in Figs. \ref{Fig:fig1}(d-f), respectively.
Figs. \ref{Fig:fig1}(a-c) show that a gap opens exactly at $E$=55 meV. The upper and lower bands do not touch together
to form a DP at $k_{y}$=0 but extra DPs may appear at $k_{y}\ne0$ \cite{MBarb2010}, see Figs. \ref{Fig:fig1}(e) and \ref{Fig:fig1}(f),
whose positions are dependent on lattice constant, and we will discuss these in details latter.
Since the center of this gap is located at zero-$\overline{k}$, we may call it zero-$\overline{k}$ gap.
One can find that the location of zero-$\overline{k}$ gap center does not shift
with lattice constant while other gaps are highly dependent on lattice constant.
As illustrated in Fig. \ref{Fig:fig2}, the position of the zero-$\overline{k}$ gap
changes with the ratio of $w_{A}/w_{B}$. It changes from $E$=80 meV, $E$=95 meV to $E$=105 meV
with the ratio $w_{A}/w_{B}$ from 1, 3/2 to 2.

The zero-$\overline{k}$ gap center could be obtained from $\overline{k}=\sum_{j=1}^{N}k_{j}w_{j}/\sum_{j=1}^{N}w_{j}$\cite{TXMa2012}.
 Using $k'$ to replace $k$ here, it is easy to find the energy for $\overline{k}$=0 as the number of berries and wells is the same.
When $w_{A}\ne w_{B}$,
\begin{eqnarray}
\lefteqn{E=\frac{2(V_{A}w_{A}^{2}-V_{B}w_{B}^{2})}{2(w_{A}^{2}-w_{B}^{2})}+\frac{t'}{2}} \nonumber\\
&&{}-\frac{\sqrt{t'^{2}(w_{A}^{2}-w_{B}^{2})^{2}+4w_{A}^{2}w_{B}^2(V_{A}-V_{B})^{2}}}{2(w_{A}^{2}-w_{B}^{2})},
\end{eqnarray}
which is different from the result in the monolayer GSLs \cite{TXMa2012,Xyafang}, due to the existence of $t'$.
If $w_{A}=w_{B}$, Eq. (6) is simplified as
\begin{equation}
E=\frac{V_{A}+V_{B}\cdot w_{B}/w_{A}}{1+w_{B}/w_{A}}+\frac{t'}{2}.
\end{equation}

In Figs. \ref{Fig:fig1}(b) and 1(c), we can also see that there is no usual DP at $k_{y}$=0
but extra DPs emerge, whose locations are determined by the trace $x_{2}$=1 \cite{Xyafang},
expending as
\begin{eqnarray}
&&x_{2}=\cos{(2q_{A}w_{A})}\cos{(2q_{B}w_{B})}}-\sin{(2q_{A}w_{A})}\sin{(2q_{B}w_{B}) \nonumber\\
&&{}\times  \frac{k_{y}^{2}(k_{A}-k_{B})^{2}+k_{A}^{2}q_{B}^{2}+k_{B}^{2}q_{A}^{2}}{2k_{A}k_{B}{q_{A}q_{B}}}=1.
\end{eqnarray}

At the normal incident angles ($k_{y}$=0), when $2q_{A}w_{A}$=$-2q_{B}w_{B}$=$m\pi$, where $m$ is an integer,
Eq. (8) always have real solutions, which means the close of zero-$\overline{k}$ gap.
For $w_{A}$=$w_{B}$=$w$, the zero-$\overline{k}$ gap shall close at
\begin{equation}
w_{m}=\frac{m\pi\hbar\upsilon _{F}}{\sqrt{(V_{A}-V_{B})^{2}-t'^{2}}},
for\quad m=1,2,3\dots.
\end{equation}

At oblique incidence ($k_{y}\ne$0), when $2q_{A}w_{A}$=$-2q_{B}w_{B}$ =$m\pi$, the extra DPs appear, which locate at
\begin{equation}
E=\frac{V_{A}+V_{B}+t'}{2}-\frac{(m\pi\hbar\upsilon_{F})^{2}}{8(V_{A}-V_{B})}\left(\frac{1}{w_{A}^{2}}-\frac{1}{w_{B}^{2}}\right),
\end{equation}
\begin{equation}
k_{y,m}=\pm\sqrt{\left(\frac{E-V_{A(B)}}{\hbar\upsilon_{F}}\right)^{2}-\left(\frac{m\pi}{2w_{A(B)}}\right)^{2}}.
\end{equation}
When $w_{A}$=$w_{B}$=$w$, Eq. (11) can be simplified as
\begin{equation}
k_{y,m}=\pm\sqrt{\frac{(V_{A}-V_{B})^{2}-t'^{2}}{4\hbar^{2}\upsilon_{F}^{2}}-\left(\frac{m\pi}{2w}\right)^{2}}.
\end{equation}
The number of extra DPs can be obtained from Eq. (12).

\begin{figure}[tbp]
\includegraphics[scale=0.45]{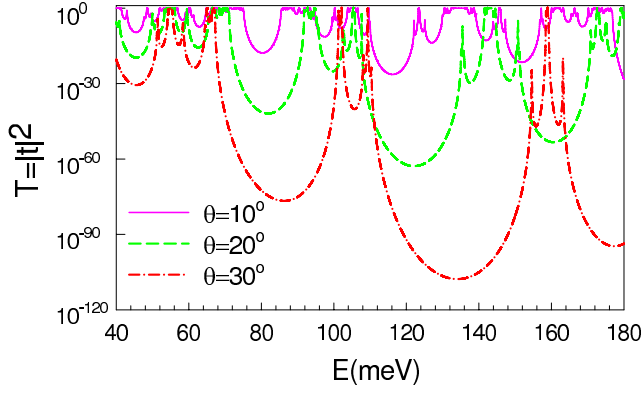}
\centering
\caption{(Color online) Dependence of transmission spectrum on different incident angles $\theta$ with fixed lattice constants $w_{A}=w_{B}=10$ nm. Other parameters are $V_{A}=150$meV, $V_{B}=0$meV, $n$=8, and $t'$=10meV.
}
\label{Fig:fig3}
\end{figure}

To learn more about the band structures, we study the effect of incident angles $\theta$
on electronic transmission spectrum. From Fig. \ref{Fig:fig3}, it is obvious that
the position of zero-$\overline{k}$ gap is weakly dependent on incident angle $\theta$ while other gaps change with $\theta$.
Moreover, one can also find that the transmission coefficient decreases
as $\theta$ increases, which means that the gaps open much wider.
\begin{figure}[tbp]
\includegraphics[scale=0.44]{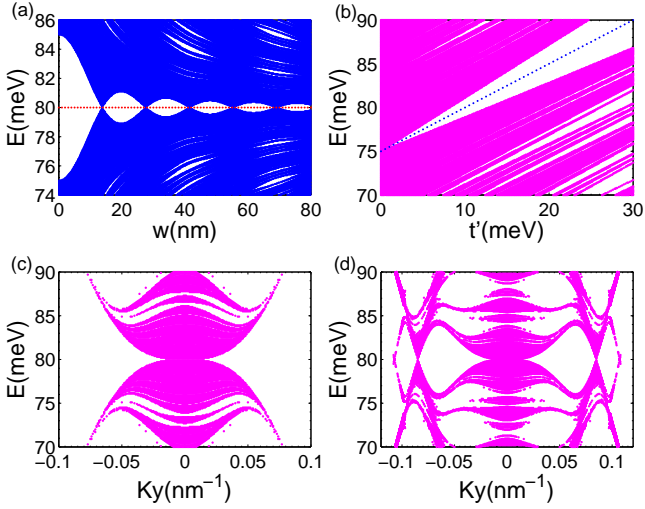}
\centering
\caption{(Color online)(a) Dependence of the band-gap structure on the lattice constant $w_{A}$=$w_{B}$=$w$ with $k_{y}$=0. (b)The effect of $t'$ on band-gap structure with $w_{A}$=$w_{B}$=20nm. Electronic band structure for BLG SLs (c) $w_{A}$=$w_{B}\approx$13.82 nm, (d) $w_{A}$=$w_{B}\approx$41.45 nm. Other parameters are $n$=8, $V_{A}$=150 meV, $V_{B}$=0 meV, and $t'$=10 meV.}
\label{Fig:fig4}
\end{figure}

Fig. \ref{Fig:fig4}(a) shows that the central position of zero-$\overline{k}$ gap is independent on lattice constant $w$ while the other gaps shift greatly. As a result of Eq. (9), the zero-$\overline{k}$ gap is gradually closed and open
oscillationally with the increasing $w$ at a period of about 13.82 nm for the used parameters.
Eq. (9) also indicates that the valence and conduction bands shall touch together periodically,  
while such touching does not lead to a DP\cite{JTwor2006,XXGuo2011},
which will be verified further in Fig. \ref{Fig:fig4}(c) and \ref{Fig:fig4}(d). It is clear that when the gap is closed, the bands touch together but they are not conical near $k_{y}$=0, while the extra Dirac points may appear at $k_{y}\ne$0, see Fig. \ref{Fig:fig4}(d).
\begin{figure}[tbp]
\includegraphics[scale=0.425]{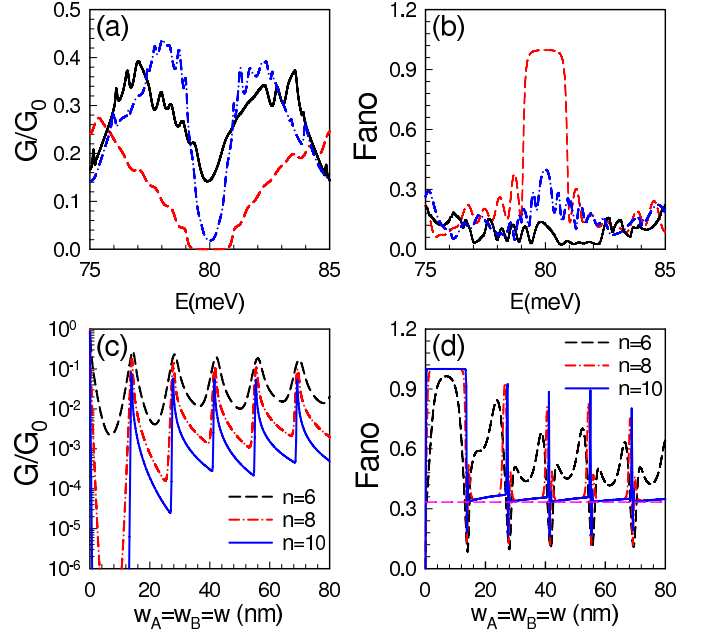}
\centering\caption{(Color online)Conductance (a) and Fano factor (b) vs Fermi energy with $n$=8. Here solid black line for $w_{A}$=$w_{B}$=13.8163 nm, dashed red line for $w_{A}$=$w_{B}$=12 nm, and dash-dotted blue line for $w_{A}$=$w_{B}$=15 nm. (c) Conductance and  (d) Fano factor are plotted vs lattice constant with energy fixed at E=80 meV. The horizonal dashed pink line denotes Fano=$1/3$ in (d). Other parameters are the same as in Fig. \ref{Fig:fig4}.}
\label{Fig:fig5}
\end{figure}
The interlayer coupling $t'$ is crucial in bilayer graphene and Fig. \ref{Fig:fig4}(b) shows
its effect on electronic band structure. The dotted blue line denotes the zero-$\overline{k}$ gap center. 
One can find that all gaps become
much wider as $t'$ increases, especially the zero-$\overline{k}$ gap.
When $t'$=0, there is no gap, while for $t'\ne$0, the zero-$\overline{k}$ gap opens.
The center of all the gaps shift versus $t'$ at a linear way
and the slopes are the same, which means that the band structure will shift as a whole.

\begin{figure}[tbp]
\includegraphics[scale=0.425]{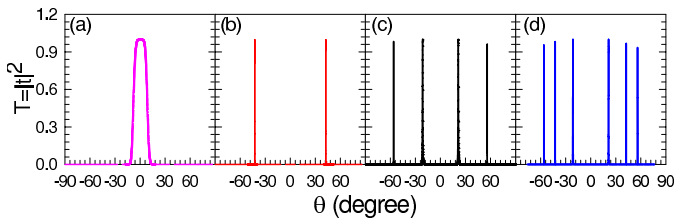}
\centering\caption{(Color online)Dependence of transmission spectrum on different incident angles with the energy fixed at $E$=80 meV are plotted. (a) $w_{A}$=$w_{B}$=13.8163 nm, (b) $w_{A}$=$w_{B}$=20 nm, (c) $w_{A}$=$w_{B}$=30 nm, (d) $w_{A}$=$w_{B}$=60 nm. Other parameters are the same as in Fig. \ref{Fig:fig4}.}
\label{Fig:fig6}
\end{figure}

Except for the transmission spectrum, other remarkable transport properties, conductance $G$ and Fano factor $F$\cite{TXMa2012,SDatta1995,JTwor2006,GF}, are demonstrated in Fig. \ref{Fig:fig5}.
Fig. \ref{Fig:fig5}(a) shows that $G$ takes its minimum with the Fermi energy in the vicinity of 80meV,
but it takes nonzero minimum value at $E$=80meV when the zero-$\overline{k}$ gap closes.
However, $G$ becomes zero with the Fermi energy in the vicinity of 80meV when the gap is open.
We should notice that $E$=80 meV is the zero-$\overline{k}$ gap center.
From Fig. \ref{Fig:fig5}(b), we can see that the Fano factor becomes
much smaller around 80 meV when the zero-$\overline{k}$ gap is closed
while it takes its maximum at $E$=80 meV when the gap exists.
The valence and conduction bands touch together when the zero-$\overline{k}$
gap is closed (In Figs. \ref{Fig:fig4}(c) and \ref{Fig:fig4}(d)),
but it is not a DP since $F$$\ne$1/3 \cite{JTwor2006,XXGuo2011}.
One can also note that $F$ takes the value of Poission process shown by the
dashed red line in Fig. \ref{Fig:fig5}(b), which can be understood by the
transmission suppression caused by the energy gap \cite{zhu2007}. 
If energy is fixed at the zero-$\overline{k}$ gap($E$=80 meV here), $G$ and Fano factor
are plotted vs the changes of lattice constant $w$ in Fig. \ref{Fig:fig5}(c)
and Fig. \ref{Fig:fig5}(d), respectively.
One can find that when there are only extra Dirac points with much larger sequence $n$,
$G$ is proportional to $\frac{1}{w-w_{m}}$, where $w_{m}$ is the critical
lattice length when the zero-$\overline{k}$ gap closes, see Eq. (9). The corresponding Fano factor is gradually approaching $1/3$.

To demonstrate the location of the extra DPs, we plot transmission
versus the incident angles of carriers in Fig. \ref{Fig:fig6} where the energy is fixed at 80 meV.
One can find that the transmission is perfect at some certain angles,
which means extra DPs appear in the band structure when $w_A=w_B >13.8163$ nm\cite{Xyafang}.
Extra DPs appear in pairs, and there are more and more pairs of extra DPs as the lattice constant increases, which can be expected from Eq. (12).
What's more, it is easy to get the location and number of extra DPs experimentally.



In summary, we have studied the electronic transport properties in the BLG TM sequence. 
The zero-$\overline{k}$ gap center is robust against the
lattice constant but sensitive to interlayer coupling $t'$.
When the extra Dirac points are generated in pairs in such structures, the electronic conductance obeys the diffusive law, $\propto \frac{1}{w-w_{m}}$, and the Fano factor
tends to 1/3, as the order of TM sequence increases.
These results provide a flexible and effective way to control the transport property in graphene, which is substantial for a lot of graphene-based electronic devices.


This work is supported by NSFCs (Grant. No. 11104014, No. 11274275, No. 61078021, and No. 11374034),
Research Fund for the Doctoral Program of Higher Education of China
20110003120007, SRF for ROCS (SEM), and the National Basic Research Program of China (Grant No. 2011CBA00108, and No. 2012CB921602 ).


\begin{thebibliography}{99}
\bibitem{Novoselov2004} K. S. Novoselov, A. K. Geim, S. V. Morozov, D.
Jiang, Y. Zhang, S. V. Dubonos, I. V. Grigorieva, and A. A. Firsov, Science {\bf 306}, 666 (2004); K. S. Novoselov, A. K. Geim, S. V. Morozov, D.
Jiang, M. I. Katsnelson, I. V. Grigorieva, S. V. Dubonos and A. A. Firsov, Nature {\bf 438}, 197(2005).

\bibitem{AHCastro2009} A. H. Castro Neto, F. Guinea, N. M. R. peters, K. s. Novoselov and A. K. Geim,  Rev. Mod. Phys {\bf 81}, 109 (2009).

\bibitem{Peres2010} N. M. R. Peres, Rev. Mod. Phys. {\bf 82}, 2673 (2010).

\bibitem{Chen2009} X. Chen and  J.-W Tao, Appl. Phys. Lett. \textbf{94},
262102 (2009).

\bibitem{Ma2010}  T. Ma, F. M. Hu, Z. B. Huang, and H.-Q. Lin, Appl. Phys. Lett. {\bf 97},
112504 (2010); F. M. Hu, T. Ma, H.-Q. Lin, and J. E. Gubernatis, Phys.
Rev. B {\bf 84}, 075414 (2011).
\bibitem{MZarenia2012} M. Zarenia, P. Vasilopoulos, and F. M. Peeters, Phys. Rev. B {\bf 85}, 245426 (2012)

\bibitem{LBrey2009} L. Brey and H. A. Fertig, Phys. Rev. Lett. {\bf 103}, 046809 (2009).

\bibitem{Zhang2009} Y. B. Zhang, T.-T. Tang, C. Girit, Z. Hao, M. C. Martin, A. Zettl, M. F. Crommie, Y. R. Shen and
F. Wang, Nature {\bf 459}, 820 (2009).

\bibitem{ALVa2008}  A. L. Vazquez de Parga, F. Calleja, B. Borca, M. C. G.
Passeggi, Jr., J. J. Hinarejos, F. Guinea, and R. Miranda, Phys. Rev. Lett. {\bf 100}, 056807 (2008).

\bibitem{JC2008}  J. C. Meyer, C. O. Girit, M. F. Crommite and A. Zettl, Appl. Phys. Lett. {\bf 92},123110  (2008).


\bibitem{Park2008} C. -H. Park, L. Yang, Y.-W. Son, M. L. Cohen, and S. G. Louie, Nat. Phys. {\bf 4}, 213 (2008).


\bibitem{PLzhao2011} P.-L. Zhao and X. Chen, Appl. Phys. Lett. {\bf 99}, 182108 (2011).
\bibitem{XXGuo2011} X.-X. Guo, D. Liu, and Y.-X. Li, Appl. Phys. Lett. {\bf 98}, 242101 (2011).
\bibitem{LGWang2010} L.-G. Wang and S.-Y. Zhu, Phys. Rev. B {\bf 81}, 205444 (2010); L.-G. Wang and X. Chen, J. Appl. Phys. {\bf 109}, 033710
(2011).
\bibitem{TXMa2012} T. Ma, C. Liang, L.-G Wang, and H.-Q. Lin, Appl. Phys. Lett. {\bf 100}, 252402 (2012).
\bibitem{Gong2012} Q. F. Zhao, J. B. Gong, and C. A. Muller, Phys. Rev. B {\bf 85}, 104201 (2012).

\bibitem{Liang2011} L. Z Tan, C.-H Park, and S. G. Louie, Nano Lett. {\bf 11}, 2596 (2011).

\bibitem{MK2011} M. Killi, S. Wu, and A. Paramekanti, Phys. Rev. Lett. {\bf 107}, 086801 (2011).

\bibitem{ZhRzh2012} Z.-R Zhang, H.-Q Li, Z.-J Gong, Y.-C Fan, T.-Q Zhang, Appl. Phys. Lett. {\bf 101}, 252104 (2012).

\bibitem{MBarb2010} M. Barbier, F. M. Peeters, and P. Vasilopoulos, Phys. Rev. B {\bf 81}, 075438 (2010).



\bibitem{Tsu2005}  R. Tsu, \textit{Superlattice to Nanoelectronics}
(Elsevier, Oxford, 2005).

\bibitem{Cott1989}  Cottam, M. G. $\&$ Tilley, D. R. \textit{Introduction to Surface and Superlattice Excitations}(Cambridge Univ.
Press, Cambridge, UK, 1989).

\bibitem{TM} TM sequence, generated by two symbols A and B, follows the inflation rules
A$\to$ AB, B$\to$ BA, and the successive TM chains are AB, ABBA, ABBABAAB$\ldots$. In Ref.\cite{TXMa2012}, some authors of us has inlustrated the way to generate TM squence in detail.


\bibitem{NLiu1997} N.-H Liu, Phys. Rev. B {\bf 55}, 3543 (1997).
\bibitem{ZCheng1988} Z. Cheng, R. Savit, and R. Merlin, Phys. Rev. B {\bf 37}, 4375 (1988).
\bibitem{Luck1989} J. M. Luck, Phys. Rev. B {\bf 39}, 5834 (1989).
\bibitem{Jiang2005} X. Y.  Jiang, Y. G. Zhang, S. L. Feng, K. C. Huang, Y. Yi, and J. D. Joannopoulos, Appl. Phys. Lett. {\bf 86}, 20111 (2005).
\bibitem{Noh2011} H. Noh, J.-K. Yang, S. V. Boriskina, M. J. Rooks, G. S. Solomon, L. D. Negro, and H. Cao, Appl. Phys. Lett. {\bf 98}, 201109 (2011).
\bibitem{Xyafang} Y. Xu, J. Zou, and G. Jin, J. Phys.: Condens. Matter {\bf 25}, 245301 (2013)
\bibitem{JTwor2006} J. Tworzydlo, B. Trauzettel, M. Titov, A. Rycerz, and C. W. J. Beenakker, Phys. Rev. Lett. {\bf 96}, 246802 (2006).
\bibitem{SDatta1995} S. Datta, \emph{Electronic Transport in Mesoscopic
Systems} (Cambridge University Press, Cambridge, England, 1995).

\bibitem{zhu2007} R. Zhu and Y. Guo, Appl. Phys. Lett. {\bf 91}, 252113 (2007).
\bibitem{GF} Here $G$ and $F$ are given by $G$=$G_{0}\int_{0}^{\pi/2}T\cos\theta_{0}d\theta_{0}$,
$F$=$\int_{-\pi/2}^{\pi/2}T(1-T)\cos\theta_{0}d\theta_{0}$/$\int_{-\pi/2}^{\pi/2}T\cos\theta_{0}d\theta_{0}$, where $T
$=$|t|^{2}$ and $G_{0}$=$2e^{2}m\upsilon _{F}L_{y}/\hbar ^{2}$,
with $L_{y}$ denoting the width of the graphene stripe in the $y$ direction.
%





\end{thebibliography}
\end{document}